\input harvmac
\input epsf

\newcount\figno
\figno=0
\def\fig#1#2#3{
\par\begingroup\parindent=0pt\leftskip=1cm\rightskip=1cm\parindent=0pt
\baselineskip=11pt
\global\advance\figno by 1
\midinsert
\epsfxsize=#3
\centerline{\epsfbox{#2}}
\vskip 12pt
\centerline{{\bf Fig. \the\figno:~~} #1}\par
\endinsert\endgroup\par
}
\def\figlabel#1{\xdef#1{\the\figno}}

\def\bosonic{1}
\def\typezero{2}
\def\bump{3}
\def\fermscatt{4}
\def\Gplot{5}

%

\overfullrule=0pt
%
%

\font\manual=manfnt \def\dbend{\lower3.5pt\hbox{\manual\char127}}

\def\ie{{\it i.e.}}
\def\eg{{\it e.g.}}
\def\cf{{\it c.f.}}

\def\sst{\scriptscriptstyle}

\def\frac#1#2{{#1\over#2}}
\def\coeff#1#2{{\textstyle{#1\over #2}}}

\def\hf{{\textstyle\half}}

\def\inbar{\,\vrule height1.5ex width.4pt depth0pt}
\def\IR{\relax{\rm I\kern-.18em R}}
\def\IC{\relax\hbox{$\inbar\kern-.3em{\rm C}$}}
\def\IQ{\relax\hbox{$\inbar\kern-.3em{\rm Q}$}}
\def\IH{\relax{\rm I\kern-.18em H}}
\def\IN{\relax{\rm I\kern-.18em N}}
\def\IP{\relax{\rm I\kern-.18em P}}
\font\cmss=cmss10
\font\cmsss=cmss10 at 7pt
\def\IZ{\relax\ifmmode\mathchoice
{\hbox{\cmss Z\kern-.4em Z}}{\hbox{\cmss Z\kern-.4em Z}}
{\lower.9pt\hbox{\cmsss Z\kern-.4em Z}}
{\lower1.2pt\hbox{\cmsss Z\kern-.4em Z}}\else{\cmss Z\kern-.4em
Z}\fi}                                                           

\catcode`\@=11
\def\slash#1{\mathord{\mathpalette\c@ncel{#1}}}
\def\underrel#1\over#2{\mathrel{\mathop{\kern\z@#1}\limits_{#2}}}

\catcode`\@=12
%

%
%
\def\ket#1{|#1\rangle}
\def\bra#1{\langle#1|}
\def\vev#1{\langle#1\rangle}

\def\sinh{{\rm sinh}} 	
\def\cosh{{\rm cosh}} 	
\def\tanh{{\rm tanh}}

\def\exp{{\rm exp}}

%
%
 \def\CA{{\cal A}}
 

 \def\CI{{\cal I}}

\def\NN{{\cal N}} 
 
\def\PP{{\cal P}} 
\def\QQ{{\cal Q}}

\def\vareps{\varepsilon}
%
%
\def\unlockat{\catcode`\@=11}
\def\lockat{\catcode`\@=12}
\unlockat
\def\newsec#1{\global\advance\secno by1\message{(\the\secno. #1)}
\global\subsecno=0\global\subsubsecno=0\eqnres@t\noindent
{\bf\the\secno. #1}
\writetoca{{\secsym} {#1}}\par\nobreak\medskip\nobreak}
\global\newcount\subsecno \global\subsecno=0
\def\subsec#1{\global\advance\subsecno
by1\message{(\secsym\the\subsecno. #1)}
\ifnum\lastpenalty>9000\else\bigbreak\fi\global\subsubsecno=0
\noindent{\it\secsym\the\subsecno. #1}
\writetoca{\string\quad {\secsym\the\subsecno.} {#1}}
\par\nobreak\medskip\nobreak}
\global\newcount\subsubsecno \global\subsubsecno=0
\def\subsubsec#1{\global\advance\subsubsecno by1
\message{(\secsym\the\subsecno.\the\subsubsecno. #1)}
\ifnum\lastpenalty>9000\else\bigbreak\fi
\noindent\quad{\secsym\the\subsecno.\the\subsubsecno.}{#1}
\writetoca{\string\qquad{\secsym\the\subsecno.\the\subsubsecno.}{#1}}
\par\nobreak\medskip\nobreak}
\def\subsubseclab#1{\DefWarn#1\xdef
#1{\noexpand\hyperref{}{subsubsection}%
{\secsym\the\subsecno.\the\subsubsecno}%
{\secsym\the\subsecno.\the\subsubsecno}}%
\writedef{#1\leftbracket#1}\wrlabeL{#1=#1}}
\lockat
%
%

%


\def\th{\theta}

\def\ep{\epsilon}
\def\vep{\varepsilon}

\def\hf{{1\over 2}}

\def\o{\over}
\def\til#1{\widetilde{#1}}

\def\bra{\langle}
\def\ket{\rangle}
\def\lf{\left}
\def\ri{\right}
\def\riya{\rightarrow}

\def\ga{\gamma}
\def\Ga{\Gamma}
\def\al{\alpha}
\def\om{\omega}

\def\rt#1{\sqrt{#1}}

\def\sitarel#1#2{\mathrel{\mathop{\kern0pt #1}\limits_{#2}}}

\def\CA{{\cal A}}

\def\CI{{\cal I}}

\def\cob{\delta}

\lref\CordesFC{
S.~Cordes, G.~W.~Moore and S.~Ramgoolam,
``Lectures on 2-d Yang-Mills theory, equivariant cohomology and topological
field theories,''
Nucl.\ Phys.\ Proc.\ Suppl.\  {\bf 41}, 184 (1995)
[arXiv:hep-th/9411210].
}

\lref\DouglasWY{
M.~R.~Douglas,
``Conformal field theory techniques in large N Yang-Mills theory,''
arXiv:hep-th/9311130.
}

\lref\PolyakovRD{
A.~M.~Polyakov,
``Quantum Geometry Of Bosonic Strings,''
Phys.\ Lett.\ B {\bf 103}, 207 (1981);
``Quantum Geometry Of Fermionic Strings,''
Phys.\ Lett.\ B {\bf 103}, 211 (1981).
}

\lref\GinspargIS{
P.~H.~Ginsparg and G.~W.~Moore,
``Lectures On 2-D Gravity And 2-D String Theory,''
arXiv:hep-th/9304011.
}

\lref\DiFrancescoNW{
P.~Di Francesco, P.~H.~Ginsparg and J.~Zinn-Justin,
``2-D Gravity and random matrices,''
Phys.\ Rept.\  {\bf 254}, 1 (1995)
[arXiv:hep-th/9306153].
}

\lref\KlebanovQA{
I.~R.~Klebanov,
``String theory in two-dimensions,''
arXiv:hep-th/9108019.
}

\lref\ShenkerUF{
S.~H.~Shenker,
``The Strength Of Nonperturbative Effects In String Theory,''
RU-90-47
{\it Presented at the Cargese Workshop on Random Surfaces, Quantum Gravity and Strings, Cargese, France, May 28 - Jun 1, 1990}
}

\lref\DavidSK{
F.~David,
``Phases Of The Large N Matrix Model And Nonperturbative Effects In 2-D
Gravity,''
Nucl.\ Phys.\ B {\bf 348}, 507 (1991).
}

\lref\McGreevyKB{
J.~McGreevy and H.~Verlinde,
``Strings from tachyons: The c = 1 matrix reloaded,''
JHEP {\bf 0312}, 054 (2003)
[arXiv:hep-th/0304224].
}

\lref\MartinecKA{
E.~J.~Martinec,
``The annular report on non-critical string theory,''
arXiv:hep-th/0305148.
}

\lref\KlebanovKM{
I.~R.~Klebanov, J.~Maldacena and N.~Seiberg,
``D-brane decay in two-dimensional string theory,''
JHEP {\bf 0307}, 045 (2003)
[arXiv:hep-th/0305159].
}

\lref\McGreevyEP{
J.~McGreevy, J.~Teschner and H.~Verlinde,
``Classical and quantum D-branes in 2D string theory,''
JHEP {\bf 0401}, 039 (2004)
[arXiv:hep-th/0305194].
}

\lref\DouglasUP{
M.~R.~Douglas, I.~R.~Klebanov, D.~Kutasov, J.~Maldacena, E.~Martinec and N.~Seiberg,
``A new hat for the c = 1 matrix model,''
arXiv:hep-th/0307195.
}

\lref\MaldacenaRE{
J.~M.~Maldacena,
``The large N limit of superconformal field theories and supergravity,''
Adv.\ Theor.\ Math.\ Phys.\  {\bf 2}, 231 (1998)
[Int.\ J.\ Theor.\ Phys.\  {\bf 38}, 1113 (1999)]
[arXiv:hep-th/9711200].
}

\lref\AharonyTI{
O.~Aharony, S.~S.~Gubser, J.~M.~Maldacena, H.~Ooguri and Y.~Oz,
``Large N field theories, string theory and gravity,''
Phys.\ Rept.\  {\bf 323}, 183 (2000)
[arXiv:hep-th/9905111].
}

\lref\DornXN{
H.~Dorn and H.~J.~Otto,
``Two and three point functions in Liouville theory,''
Nucl.\ Phys.\ B {\bf 429}, 375 (1994)
[arXiv:hep-th/9403141].
}

\lref\ZamolodchikovAA{
A.~B.~Zamolodchikov and A.~B.~Zamolodchikov,
``Structure constants and conformal bootstrap in Liouville field theory,''
Nucl.\ Phys.\ B {\bf 477}, 577 (1996)
[arXiv:hep-th/9506136].
}

\lref\FateevIK{
V.~Fateev, A.~B.~Zamolodchikov and A.~B.~Zamolodchikov,
``Boundary Liouville field theory. I: Boundary state and boundary  two-point
function,''
arXiv:hep-th/0001012.
}

\lref\TeschnerMD{
J.~Teschner,
``Remarks on Liouville theory with boundary,''
arXiv:hep-th/0009138.
}

\lref\ZamolodchikovAH{
A.~B.~Zamolodchikov and A.~B.~Zamolodchikov,
``Liouville field theory on a pseudosphere,''
arXiv:hep-th/0101152.
}

\lref\TeschnerRV{
J.~Teschner,
``Liouville theory revisited,''
Class.\ Quant.\ Grav.\  {\bf 18}, R153 (2001)
[arXiv:hep-th/0104158].
}

\lref\WittenYR{
E.~Witten,
``On string theory and black holes,''
Phys.\ Rev.\ D {\bf 44}, 314 (1991).
}

\lref\MandalTZ{
G.~Mandal, A.~M.~Sengupta and S.~R.~Wadia,
``Classical solutions of two-dimensional string theory,''
Mod.\ Phys.\ Lett.\ A {\bf 6}, 1685 (1991).
}

\lref\PolchinskiMB{
J.~Polchinski,
``What is string theory?,''
arXiv:hep-th/9411028.
}

\lref\TakayanagiSM{
T.~Takayanagi and N.~Toumbas,
``A matrix model dual of type 0B string theory in two dimensions,''
JHEP {\bf 0307}, 064 (2003)
[arXiv:hep-th/0307083].
}

\lref\DeWolfeQF{
O.~DeWolfe, R.~Roiban, M.~Spradlin, A.~Volovich and J.~Walcher,
``On the S-matrix of type 0 string theory,''
JHEP {\bf 0311}, 012 (2003)
[arXiv:hep-th/0309148].
}

\lref\KapustinHI{
A.~Kapustin,
``Noncritical superstrings in a Ramond-Ramond background,''
JHEP {\bf 0406}, 024 (2004)
[arXiv:hep-th/0308119].
}

\lref\BerkovitsTG{
N.~Berkovits, S.~Gukov and B.~C.~Vallilo,
``Superstrings in 2D backgrounds with R-R flux and new extremal black  holes,''
Nucl.\ Phys.\ B {\bf 614}, 195 (2001)
[arXiv:hep-th/0107140].
}

\lref\GukovYP{
S.~Gukov, T.~Takayanagi and N.~Toumbas,
``Flux backgrounds in 2D string theory,''
JHEP {\bf 0403}, 017 (2004)
[arXiv:hep-th/0312208].
}

\lref\DavisXB{
J.~Davis, L.~A.~Pando Zayas and D.~Vaman,
``On black hole thermodynamics of 2-D type 0A,''
JHEP {\bf 0403}, 007 (2004)
[arXiv:hep-th/0402152].
}

\lref\DanielssonXF{
U.~H.~Danielsson, J.~P.~Gregory, M.~E.~Olsson, P.~Rajan and M.~Vonk,
``Type 0A 2D black hole thermodynamics and the deformed matrix model,''
JHEP {\bf 0404}, 065 (2004)
[arXiv:hep-th/0402192].
}

\lref\WittenZW{
E.~Witten,
``Anti-de Sitter space, thermal phase transition, and confinement in  gauge
theories,''
Adv.\ Theor.\ Math.\ Phys.\  {\bf 2}, 505 (1998)
[arXiv:hep-th/9803131].
}

\lref\SusskindVY{
L.~Susskind,
``Matrix theory black holes and the Gross Witten transition,''
{\it Prepared for ICTP Conference on Super Five Brane Physics in 5+1 Dimensions, Trieste, Italy, 1-3 Apr 1998}
}

\lref\MooreZV{
G.~W.~Moore, M.~R.~Plesser and S.~Ramgoolam,
``Exact S matrix for 2-D string theory,''
Nucl.\ Phys.\ B {\bf 377}, 143 (1992)
[arXiv:hep-th/9111035].
}

\lref\MaldacenaIX{
J.~M.~Maldacena and A.~Strominger,
``Black hole greybody factors and D-brane spectroscopy,''
Phys.\ Rev.\ D {\bf 55}, 861 (1997)
[arXiv:hep-th/9609026].
}

\lref\BirrellIX{
N.~D.~Birrell and P.~C.~W.~Davies,
``Quantum Fields In Curved Space,''
}

\lref\DiFrancescoUD{
P.~Di Francesco and D.~Kutasov,
``World sheet and space-time physics in two-dimensional (Super)string theory,''
Nucl.\ Phys.\ B {\bf 375}, 119 (1992)
[arXiv:hep-th/9109005].
}

\lref\GiveonPX{
A.~Giveon and D.~Kutasov,
``Little string theory in a double scaling limit,''
JHEP {\bf 9910}, 034 (1999)
[arXiv:hep-th/9909110].
}

\lref\HoriAX{
K.~Hori and A.~Kapustin,
``Duality of the fermionic 2d black hole and N = 2 Liouville theory as  mirror
symmetry,''
JHEP {\bf 0108}, 045 (2001)
[arXiv:hep-th/0104202].
}

\lref\KazakovPM{
V.~Kazakov, I.~K.~Kostov and D.~Kutasov,
``A matrix model for the two-dimensional black hole,''
Nucl.\ Phys.\ B {\bf 622}, 141 (2002)
[arXiv:hep-th/0101011].
}

\lref\MarcusFP{
N.~Marcus and Y.~Oz,
``The Spectrum of the 2-D black hole or does the 2-D black hole have tachyonic
or W hair?,''
Nucl.\ Phys.\ B {\bf 407}, 429 (1993)
[arXiv:hep-th/9305003].
}

\lref\WittenFP{
E.~Witten,
``Two-dimensional string theory and black holes,''
arXiv:hep-th/9206069.
}

\lref\senparis{
A.~Sen,
``Symmetries and Conserved Charges in 2d String Theory,''
Talk at Strings2004, Paris (June 28-July 2, 2004).
}

\lref\DemeterfiCM{
K.~Demeterfi, I.~R.~Klebanov and J.~P.~Rodrigues,
``The Exact S matrix of the deformed c = 1 matrix model,''
Phys.\ Rev.\ Lett.\  {\bf 71}, 3409 (1993)
[arXiv:hep-th/9308036].
}

\lref\AharonyXN{
O.~Aharony, A.~Giveon and D.~Kutasov,
``LSZ in LST,''
arXiv:hep-th/0404016.
}

\lref\GrossMD{
D.~J.~Gross and I.~R.~Klebanov,
``Vortices And The Nonsinglet Sector Of The C = 1 Matrix Model,''
Nucl.\ Phys.\ B {\bf 354}, 459 (1991).
}

\lref\CallanRS{
C.~G.~.~Callan, S.~B.~Giddings, J.~A.~Harvey and A.~Strominger,
``Evanescent black holes,''
Phys.\ Rev.\ D {\bf 45}, 1005 (1992)
[arXiv:hep-th/9111056].
}


\def\alp{\alpha'}
\def\H{{\sst\rm H}}
\def\adm{{\sst\rm ADM}}

\Title{
\vbox{\hbox{EFI-04-28}
      \hbox{hep-th/0407136}}
}
{Scattered Results in 2D String Theory}

\vskip .2in

\centerline{Emil Martinec and Kazumi Okuyama}

\vskip .2in

\centerline{Enrico Fermi Institute, University of Chicago} 
\centerline{5640 S. Ellis Ave., Chicago IL 60637, USA}
\centerline{\tt ejm@theory.uchicago.edu, kazumi@theory.uchicago.edu}

\vskip 3cm
\noindent

\noindent
The nonperturbative $1\to N$ tachyon scattering amplitude
in 2D type 0A string theory is computed.
The probability that $N$ particles are produced
is a monotonically decreasing function of $N$
whenever $N$ is large enough that statistical methods apply.
The results are compared with expectations from black hole
thermodynamics.

\Date{July 2004}
\vfil\eject


\newsec{Introduction}

String theory outside the critical dimension
has proven a fruitful subject of investigation, yielding
many valuable insights.  The introduction of such
``non-critical'' strings by Polyakov 
\PolyakovRD\
highlighted the importance of local scale invariance
as a consistency condition on the worldsheet QFT
describing string propagation.
Indeed, noncritical string models are now understood
as a particular class of backgrounds of ordinary,
``critical'' string theory with asymptotically linear dilaton:
\eqn\lindil{
  G_{\mu\nu}\sim \eta_{\mu\nu}\quad,\qquad \Phi\sim n\cdot X
}
where $n^2=\frac{26-D}{6\alp}$ for the bosonic string,
and $n^2=\frac{10-D}{4\alp}$ for the fermionic string.
Perturbative string theory in such a background is self-consistent
if the dynamics avoids the region of strong coupling.
One way to accomplish this (for $D\le 2$)
is to turn on a closed string tachyon background
\eqn\liouwall{
  T = \mu \, e^{\alpha n\cdot X}\ .
}
The resulting worldsheet potential pushes strings away from
strong coupling.

The formulation of non-critical string models in target
dimension $D\le 2$ in terms of integrals over random matrices
\refs{\DiFrancescoNW,
\GinspargIS,
\KlebanovQA}
provided the first insights into stringy non-perturbative effects
\refs{\DavidSK,\ShenkerUF}.
Indeed, recent work 
\refs{\McGreevyKB,\MartinecKA,\KlebanovKM,\McGreevyEP,%
\TakayanagiSM,\DouglasUP}
has established that the matrix model describes an open string
dual of non-critical string theory background
\lindil-\liouwall, in which the matrix
eigenvalues are D-branes of the non-critical string.

Thus these string backgrounds provide particularly tractable
examples of the correspondence between gauge theory
and gravity uncovered in recent years
\MaldacenaRE\ (for a review, see \AharonyTI).
Here, both sides of the correspondence are amenable
to precise calculation.  Namely, on the 
matrix model/open string/gauge theory side, the matrix path integral
is exactly solvable.  On the closed string/gravity side,
the correlation functions of the worldsheet CFT of the
non-critical string can be computed exactly using
conformal bootstrap methods
\refs{\DornXN,
\ZamolodchikovAA,
\FateevIK,
\TeschnerMD,
\ZamolodchikovAH}
(for a review, see \TeschnerRV).
The agreement between the two approaches
of all quantities thus far computed 
provides strong evidence that the two formulations
are equivalent.

On the closed string side, another interesting non-critical string
background is the 2D black hole
\refs{\WittenYR,\MandalTZ}
\eqn\twodbh{
\eqalign{
  ds^2 &= \frac k{\alp}\bigl[dr^2 - \tanh^2 r\, dt^2 \bigr]\cr
  \Phi &= \Phi_0 - 2 \log[\cosh \,r]
}}
with $k=9/4$ for the bosonic string,
and $k=5/2$ for the fermionic string.
This geometry is in the class of asymptotically linear 
dilaton backgrounds \lindil, and the Euclidean geometry
admits an exact worldsheet CFT description as 
the gauged WZW model $SL(2,\IR)/U(1)$.
These results raise the possibility of a non-perturbative
and perhaps solvable matrix model formulation of 2D black hole dynamics.

An obstacle to the realization of this idea in the bosonic string
is the non-perturbative ambiguity in the definition of
the corresponding matrix model.  In the latter, the matrix
eigenvalues are free fermions in an inverted quadratic potential,
with only the metastable states on one side of the
barrier filled (see figure \bosonic a).  
The interpretation of eigenvalues
in the ``ground state'' tunneling to the other side of the barrier,
or simply vaulting over the barrier in a high energy process
as in figure \bosonic b,
remained unclear.  Since black hole formation is expected
to arise through such high energy processes
(\cf\ \PolchinskiMB\ for a discussion in the present context),
the status of black holes in the matrix model also remained unclear.

\bigskip
{\vbox{{\epsfxsize=4.7in
        \nobreak
    \centerline{\epsfbox{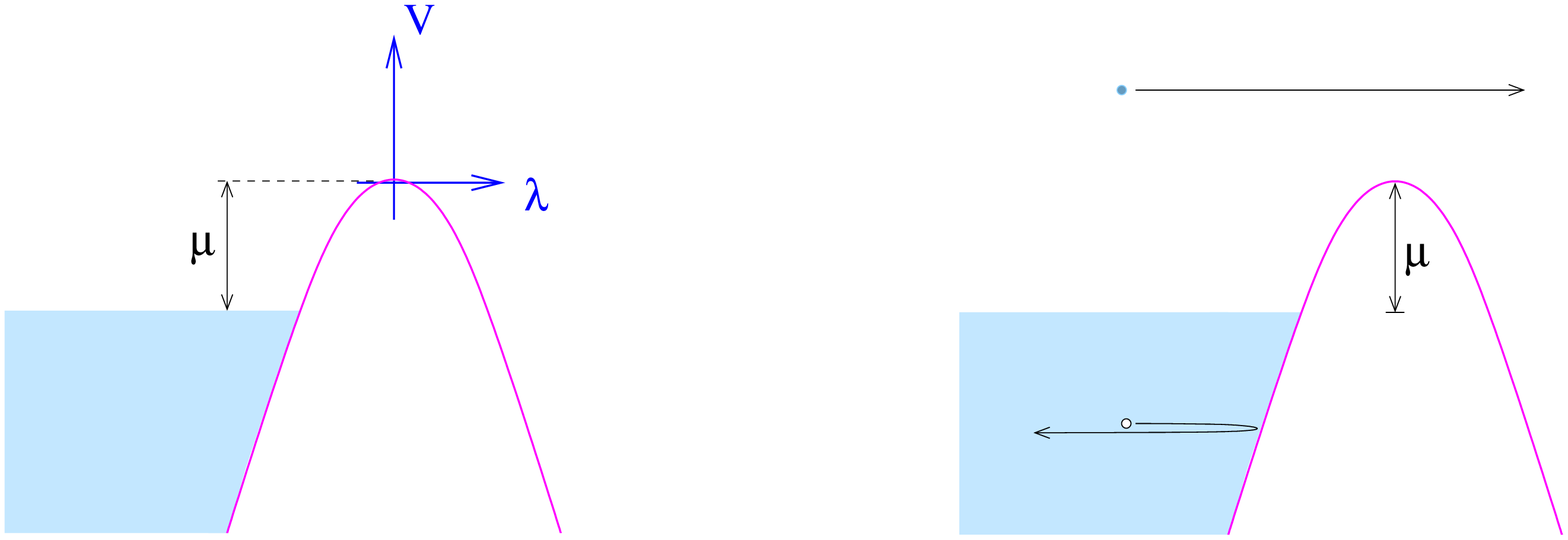}}
        \nobreak\bigskip
    {\raggedright\it \vbox{
{\bf Figure \bosonic.}
{\it
The matrix model for the bosonic string:
(a) The metastable ``ground state'';
(b) The puzzle introduced by high energy scattering.
} }}}}
\bigskip}

Now it is understood that the stable ground state,
where eigenvalue fermions fill both sides of the harmonic barrier,
is a matrix model formulation of the superstring --
specifically, its type 0B incarnation
\refs{\TakayanagiSM,\DouglasUP} (see figure \typezero a).
It seems appropriate to revisit the question of black hole
formation in the matrix model.

\bigskip
{\vbox{{\epsfxsize=4.5in
        \nobreak
    \centerline{\epsfbox{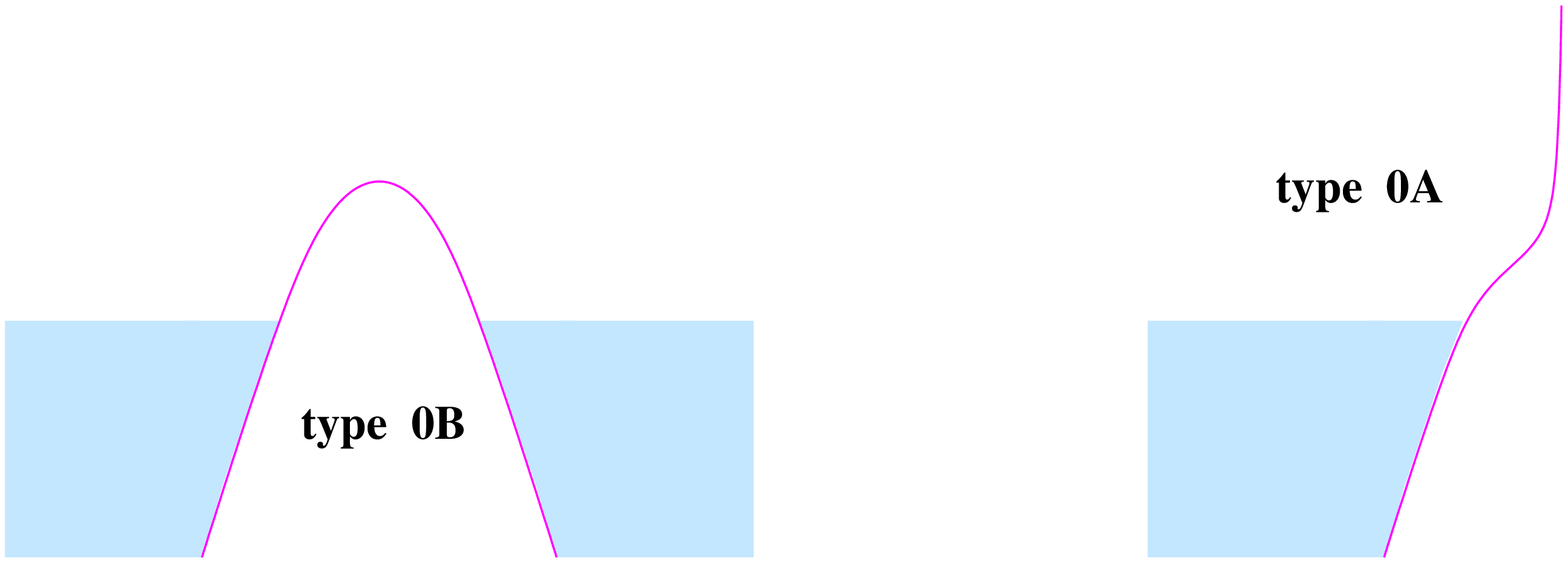}}
        \nobreak\bigskip
    {\raggedright\it \vbox{
{\bf Figure \typezero.}
{\it
The matrix model eigenvalue potentials 
and ground state configurations for the 
(a) type 0B, and (b) type 0A fermionic string.
} }}}}
\bigskip}

For reasons of technical simplification, we will consider
the related type 0A matrix model, which allows us
to sidestep various subtleties associated to the
intercommunication of the two sides of the harmonic barrier
in the 0B model
\DeWolfeQF.
The eigenvalue potential of the 0A model \refs{\DouglasUP,\KapustinHI}
\eqn\zeroapotl{
  V = -\frac{\lambda^2}{4\alp} + \frac{q^2-\coeff14}{2\lambda^2}\ ,
}
depicted in figure \typezero b, has
eigenvalues restricted to the half-line $\lambda<0$.
The parameter $q$ is the RR charge of the background.
For $q=0$ the black hole solution of \twodbh\ applies
(with $k=5/2$), 
and can again be described on the
worldsheet as the $\NN=1$ supersymmetric $SL(2,\IR)/U(1)$
coset model. The more general black holes
with $q\ne 0$ have been studied in 
\refs{\BerkovitsTG,\GukovYP,\DavisXB,\DanielssonXF}.

Can one form black holes in 2D string theory?
In any situation where black holes form, one expects them
to constitute the generic intermediate states due to their
large density of states, and therefore to 
dominate the behavior of the S-matrix.  Thus we would like to ask
whether the S-matrix of 2D string theory has any characteristics
that one would associate with the appearance of black
holes as intermediate states.  In higher dimensions,
this is indeed the case in semiclassical gravity;
the generic result of a very energetic collision
is the formation of a black hole.

Nevertheless, two arguments suggest that black holes do not form
in 2D string theory.
\item{1.}
First, the matrix model has infinitely many conserved 
quantities, since the underlying dynamical objects
are free fermions.  Any black hole will correspond
to rather specific values of these conserved quantities,
and therefore the generic initial state having other
values for these quantities will have small overlap 
with the black hole.
\item{2.}
In other examples of the gauge/gravity correspondence,
the formation of black holes is associated
with a deconfinement transition in the gauge theory,
in which one accesses non-singlet degrees of freedom
in the gauge theory
\WittenZW.
These degrees of freedom are projected out of the
matrix model for 2D string theory, suggesting
that there is no regime in which the dominant
configurations in the dynamics can be associated
to black holes.

\noindent
These two issues are related.  The appearance
of free fermions and their associated hierarchy of conserved
quantities is directly related to the absence of non-singlet
degrees of freedom in the matrix model, which would be
needed to provide the exponentially large degeneracy
of black hole states.  
The projection onto singlets is consistent with the Gauss law 
of the gauge theory on D0-branes, 
which provide the worldsheet description of
matrix eigenvalues pulled out of the Fermi sea 
\DouglasUP.%
\foot{Even if non-singlet states were somehow allowed, 
their energy cost becomes infinite 
in the continuum limit of the matrix model
\GrossMD.}

Even so, it would be nice to verify this conventional
wisdom with a concrete calculation, to see if there are
any indications of an object with the
characteristics of a black hole which participates
as an intermediate state in the dynamics.  
For example, it might be possible for a forming black hole
to shed via tachyon radiation
the collection of conserved charges along with some
fraction of the infall energy, leaving behind a
long-lived black hole intermediate state.

Below we will examine
high energy scattering processes in the matrix model,
using the formalism of \MooreZV\ for the exact S-matrix.
Of course, the high energy behavior of the S-matrix is
of intrinsic interest, independent of the question of
whether black holes are involved in the scattering.
We wish to look for the probability to find $N$ particles
in the final state, to see if there is a correspondence
with the expected decay spectrum of a 2D black hole.
We will examine in detail the $1\to N$ scattering amplitude,
but we expect other amplitudes involving a finite number
of high energy incoming particles to behave similarly.
Is there any feature in the outgoing spectrum
that we might identify with a black hole intermediate state?
We will show that, for $N$ more than a few, 
the probability to produce $N$ particles is a rapidly decreasing
function of $N$, with no feature in the range $N\sim \omega$
that would be characteristic of black hole intermediate states.


\newsec{The black hole in type 0A gravity}

Before embarking on an analysis of scattering in the
matrix model, let us consider what result would be
predicted by the appearance of 2D black holes as
the generic intermediate state. 

At lowest order in $\alp$, the effective action of type 0A
string theory admits a class of RR charged black hole solutions
\refs{\BerkovitsTG,
\GukovYP,
\DavisXB,
\DanielssonXF}.
In Schwarzchild-like coordinates, the geometry can be written
(for $\mu=0$, \ie\ no tachyon field background)
\eqn\zeroasol{\eqalign{
  ds^2 &= -l(\phi)dt^2+{d\phi^2\o l(\phi)} \cr
  \Phi &= \kappa\phi
}}
where
$\kappa=\rt{2\o\al'}$, 
and $l(\phi)$ is given by
\eqn\lphi{
  l(\phi)=1+e^{2\kappa(\phi-\phi_\H)}
	\Big[2\kappa e^{-\vep}(\phi-\phi_\H)-1\Big]\ .
}
Here, $\phi=\phi_{\H}$ is the location of the black hole horizon;
the horizon of the extremal black hole is located at
\eqn\phiext{
  \phi_e=\frac{1}{2\kappa}\log\Bigl[{32\pi\o q^2}\Bigr]\ ,
}
and
\eqn\phiHep{
\vep=2\kappa(\phi_e-\phi_{\H})
}
is a measure of the departure from extremality.


In terms of these quantities, the 
Hawking temperature and ADM mass above extremality are given by
\eqn\Tmass{\eqalign{
 T &= {\kappa\o2\pi}(1-e^{-\vep}) 	\cr
 E_{\adm} &= 2\kappa e^{-2\kappa\phi_e}\Big[e^{\vep}-1-\vep\Big]\ .
}}
Note that the temperature rises smoothly from zero
at extremality to a limiting temperature of order one in string units.
The high mass behavior is thus Hagedorn thermodynamics,
with an entropy
\eqn\hagedorn{
  S_{\sst\rm BH} \sim \beta_\H M
}
where $\beta_\H = 2\pi\sqrt{\alp/2}$.

The Hawking temperature is related to the actual spectrum
of radiation seen at infinity via so-called greybody factors 
(\cf\ \BirrellIX).
The potential seen by a Hawking quantum as it travels
away from the horizon acts as a filter and distorts
the spectrum.  Let us estimate the magnitude
of such effects.

The equation of motion for the closed string tachyon 
$\til{T}=e^{-\Phi}T$ in this background is 
\eqn\eomT{
  \lf(l{d\o d\phi}\ri)^2\til{T}+(\om^2-V)\til{T}=0\ ,
}
where
\eqn\potV{\eqalign{
  V &= 7\kappa ll'-15\kappa^2l(l-1)	\cr
    &= \kappa^2le^{2\kappa(\phi-\phi_\H)}
	\Big[1+14e^{-\vep}-2\kappa e^{-\vep}(\phi-\phi_\H)\Big]
}}
A sketch of this effective potential is shown in figure \bump.

\bigskip
{\vbox{{\epsfxsize=3in
        \nobreak
    \centerline{\epsfbox{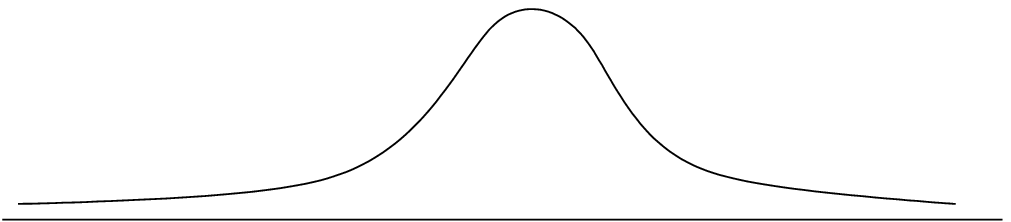}}
        \nobreak\bigskip
    {\raggedright\it \vbox{
{\bf Figure \bump.}
{\it
The effective potential seen by a tachyon perturbation
in the 0A black hole background, as a function of
the ``tortoise coordinate'' $x=\int l^{-1}d\phi$.
} }}}}
\bigskip}

The height of the effective potential is of order one
in string units.  The transmission amplitude rises
to unity for incident energies far above the barrier height,
and vanishes smoothly as the incident energy goes to zero.
At any rate, for black holes of mass 
above extremality $E_\adm\gg \kappa$,
where the Hawking temperature is of order $T\sim \kappa/2\pi$,
the greybody effects amount to a factor of order one.
It is only for near-extremal black holes, where the typical
Hawking quantum has an energy small compared to the barrier
height, that greybody effects will be significant.

One can also consider black hole formation in the context
of the low-energy effective field theory.
The gravitational back-reaction of an infalling tachyon pulse
propagating according to \eomT\ naively results in black hole
formation for any sufficiently energetic pulse.%
\foot{As in the study of dilaton gravity 
coupled to conformal matter \CallanRS,
we would have to consider a incoming tachyon pulse localized along $\CI^-$
in order to have a finite energy density which is
sufficiently localized to create a black hole.
This can be achieved by making a tachyon wave packet 
instead of the plane wave state which
we will consider in the following sections.
However, this does not change the qualitative feature of 
the number distribution in the final state discussed in section 5 
(see also footnote 4).
We would like to thank A. Strominger for raising this issue.}
The horizon moves out to large $\phi$, and therefore
weak coupling, with increasing energy.
The bulk of the pulse is not backscattered by
the effective potential $V$, equation \potV.
Provided that $\alp$ corrections to the low-energy
field equations do not change the qualitative
structure deduced from the leading order equations, 
the black hole is the generic
intermediate state for any highly energetic initial state --
once tachyons fall inside a trapped surface,
the black hole will form.  The trapped surface forms
at weak coupling for sufficiently large energy, and
so the classical field equations are reliable.

The qualitative picture that then emerges is as follows:
Energetic incoming states form black holes.
For black hole masses far above extremality,
$E_\adm\gg \kappa q^2$, the Hawking temperature $T_\H$ is of order one
in string units.  The decay of the black hole will proceed
smoothly via the emission of quanta whose energy is of
order one in string units, until $E_\adm\sim \kappa q^2$.
At this point the character of the decay changes; 
the temperature varies with the mass, smoothly
decreasing with as the black hole evaporates toward
extremality.  We can make these endpoint effects parametrically
small by considering black holes whose initial mass
is large compared with the charge.  
The expectation then is that the typical number 
of quanta emitted in a black hole
decay will be of order the mass in string units,
\eqn\expnum{
  \vev{N} \sim \frac{E_\adm}{T_{\H}} = \frac{E_\adm}{\kappa/2\pi}\ .
}

It should however be emphasized that the geometry \zeroasol-\lphi\
is strongly curved in string units, and the above picture
of its properties could be strongly corrected by stringy effects.
It is somewhat reassuring in this regard,
that for $q=0$ the black hole
admits an exact CFT description as an $SL(2,\IR)/U(1)$
gauged WZW model.  Even so, this may also be misleading,
as we will discuss in the final section.

\newsec{The type 0A S-matrix}

The asymptotic states of type 0A string theory are 
perturbations of the tachyon field, realized in the matrix
description as disturbances of the eigenvalue density.
To evaluate the S-matrix, 
one is instructed to perform a kind of LSZ reduction
on eigenvalue density correlators, taking the point of 
evaluation of the corresponding fermion bilinears to
$|\lambda|\to\infty$, $t\to\pm\infty$ and extracting
the coefficient of the asymptotic behavior.
The connected correlation function of the asymptotic eigenvalue density
perturbations consists of ring diagrams of the corresponding
fermion bilinears; an example of $1\to N$ scattering is
depicted in figure \fermscatt.

\bigskip
{\vbox{{\epsfxsize=2.5in
        \nobreak
    \centerline{\epsfbox{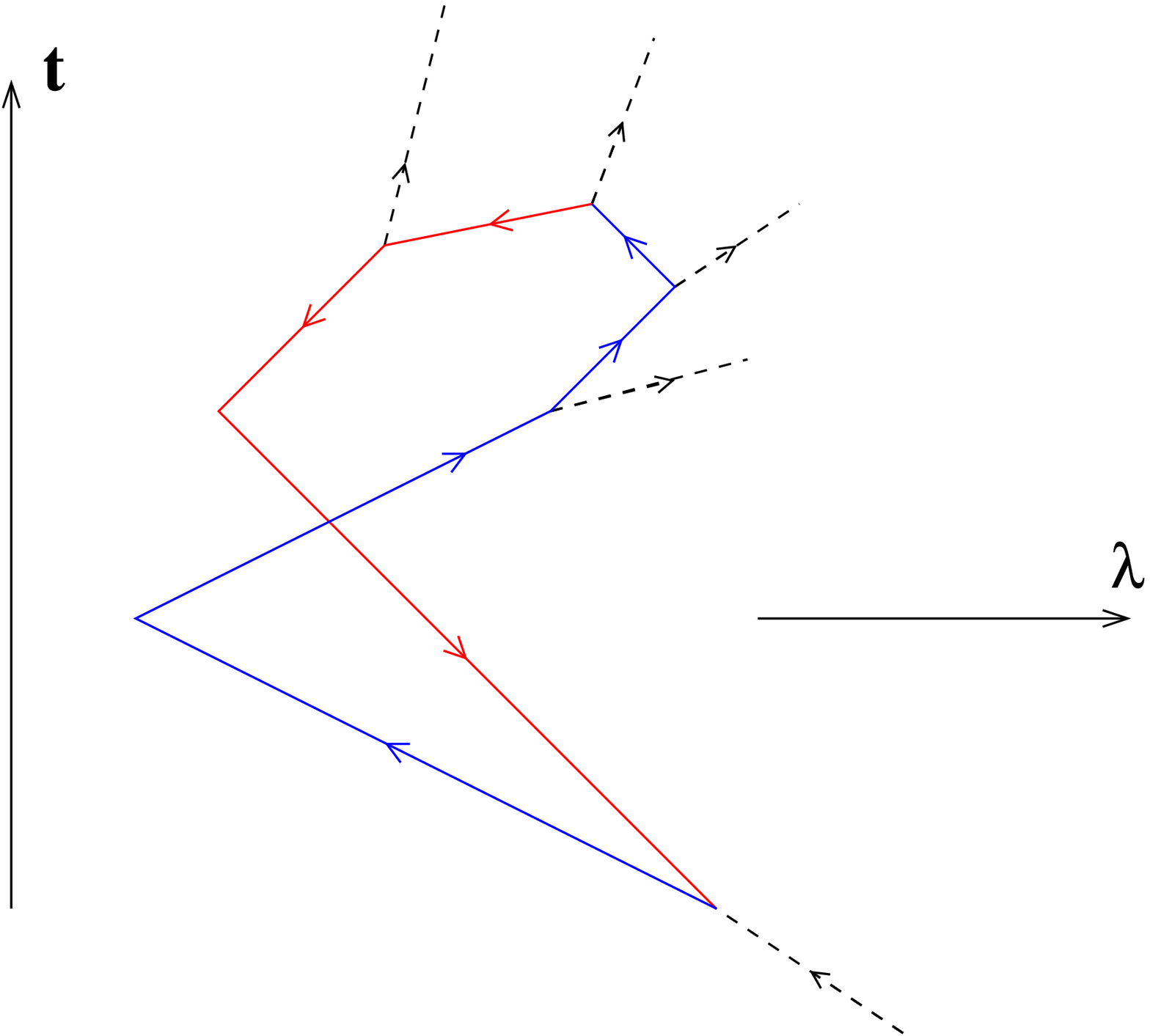}}
        \nobreak\bigskip
    {\raggedright\it \vbox{
{\bf Figure \fermscatt.}
{\it
$1\to N$ scattering in the matrix model.
The incoming tachyon (dashed line) splits into a particle (blue) and
hole (red), which reflect off the potential wall.
The particle-hole pair decays back toward the Fermi surface
via the emission of final state tachyons.
} }}}}
\bigskip}

A recipe for the result of this calculation was given in 
\MooreZV\ (and reviewed in \GinspargIS).
One picks from the products of fermion propagators
the terms in which the wildly oscillating phases
$\exp[\frac i2\lambda^2]$ cancel; one then extracts the
coefficients of the asymptotic behavior of the density field
\eqn\asdens{
  \rho(\lambda,t)\sim \frac{1}{\lambda}
	\,\exp[-i\omega(t\pm\log|\lambda|)]
}
to obtain the scattering matrix element.
For each fermion reflecting off the potential
wall near the origin in $\lambda$ space, there is a factor
of the reflection coefficient $R(\omega)$ for particles
and $R^*(\omega)$ for holes.  
The reflection coefficient of type 0A is given by%
\foot{We set $\al'=2$.}
\eqn\Rom{
R(\om)=\lf|{q^2\o4}-{1\o16}+\mu^2\ri|^{-i\om}
{\Ga(\hf+{q\o2}+i\om-i\mu)\o\Ga(\hf+{q\o2}-i\om+i\mu)}
}

In comparing to worldsheet results, there are in addition
certain ``leg-pole factors'' relating the matrix model
density perturbation to the string tachyon
\DiFrancescoUD.
These are energy-dependent phases for on-shell scattering processes,
which therefore affect quantities such as the time delay
of an outgoing particle.  However, they will not affect the
outgoing particle distribution as a function of energy, 
which is the object we study here.  Therefore we
henceforth ignore these factors.%
\foot{The leg-pole factors could potentially alter the result
when considering a wave packet of incoming tachyon. 
The width $\delta\omega$ of the wavepacket determines the localization in
position space.  One can choose the width 
much smaller than the string mass, so that
the leg pole factor is constant over the width
of the wavepacket in frequency space, and yet much larger than the
inverse of the time scale $T$ that the entire wavepacket spends
in the strong coupling region, behind the point in $\phi$ where the low
energy field equations would predict that a horizon forms.  
The variation of the amplitude is smooth over the
width of such a wavepacket, 
and so the result of scattering such a localized
wavepacket will not differ substantially
from that of a pure plane wave.}

Our operating hypothesis, as explained in the last section,
is that any sufficiently energetic incoming state 
causes the formation of a black hole.
We will thus concentrate on the $1\to N$ amplitude for simplicity
(a partial analysis of the $2\to N$ amplitude is contained in
the Appendix). 
The momenta of the external states then divide into two groups:
The set $S$ that attach to the propagating fermion line, 
and therefore have positive frequency;
and the set $\overline S$ that attach to the propagating hole line,%
\foot{There is always one tachyon emitted at the point where
the particle and hole lines meet in the final state;
in our conventions, this tachyon belongs to the set $\overline S$.}
and therefore have negative frequency.
According to the prescription outlined above, 
the scattering amplitude is 
\eqn\AonetoN{
{\cal A}_N(\om|\om_i)=\sum_{S\subset\{1,\cdots, N\}}(-1)^{|S|}
\int_{\om_S}^\om d\xi\, R(\om-\xi)R^*(-\xi)
}
where $\om_S=\sum_{l=1}^k\om_{j_l}$ and $|S|=k$
for $S=\{j_1,\cdots,j_k\}$.

Plugging the explicit form of reflection coefficient \Rom\ into 
\AonetoN, ${\cal A}_N$ is rewritten as
\eqn\AasFint{
{\cal A}_N=\lf|{q^2\o4}-{1\o16}+\mu^2\ri|^{-i\om}
\sum_{S}(-1)^{|S|}\int_{\om_S-\mu}^{\om-\mu}d\xi F(\om-\xi)F(\xi)
}
where we defined $F(\xi)$ by
\eqn\Fxidef{
F(\xi)={\Ga(\hf+{q\o2}+i\xi)\o\Ga(\hf+{q\o2}-i\xi)}.
}
Dropping the overall phase factor $|{q^2\o4}-{1\o16}+\mu^2|^{-i\om}$
and changing the integration variable as $\xi\riya {\xi+\om\o2}$,
the amplitude ${\CA}_N$ in \AasFint\ becomes
\eqn\tilANshift{
  {\CA}_N
  =\hf\sum_{S}(-1)^{|S|}\int_{\om_S-\om_{\bar{S}}-2\mu}^{\om-2\mu}d\xi 
	F\lf({\om-\xi\o2}\ri)F\lf({\om+\xi\o2}\ri).
}

In \MooreZV, a saddle point approximation was applied directly to
this expression \tilANshift\ of amplitude.
However, it is more useful to perform a summation over $S$
before doing the saddle point approximation.  
For this purpose, 
we utilize the trick of inserting ``1'' into the integral:
\eqn\onecob{
1=\int_{-\infty}^\infty d\nu\cob(\nu-\xi)=\int_{-\infty}^\infty
{d\nu dt\o2\pi}e^{it(\nu-\xi)}.
}
Then \Fxidef\ becomes
\eqn\xiint{\eqalign{
\int_{\om_S-\om_{\bar{S}}-2\mu}^{\om-2\mu}d\xi f(\xi)&=
\int_{\om_S-\om_{\bar{S}}-2\mu}^{\om-2\mu}d\xi\int_{-\infty}^\infty
{d\nu dt\o2\pi}e^{it(\nu-\xi)}f(\nu) \cr
&=\int_{-\infty}^\infty
{d\nu dt\o2\pi}e^{it(\nu+2\mu)}f(\nu){1\o -it}\Big(e^{-it\om}-
e^{it(\om_{\bar{S}}-\om_S)}\Big)
}}
where we defined $f(\xi)=F({\om-\xi\o2})F({\om+\xi\o2})$.
Now we can perform the summation over $S$
by using the relations $\sum_S(-1)^{|S|}=0$ and
\eqn\sumSprod{
\sum_S(-1)^{|S|}e^{it(\om_{\bar{S}}-\om_S)}
=\prod_{j=1}^N(e^{it\om_j}-e^{-it\om_j}).
}
Finally, we arrive at a compact form of ${\cal A}_N$
\eqn\ANfinalform{
{\cal A}_N=i^{N-1}\int_{-\infty}^{\infty}{dt\o2t}G(t)e^{2it\mu}
\prod_{j=1}^N2\sin\om_jt
}
where $G(t)$ is given by
\eqn\GtFourier{
G(t)=\int_{-\infty}^\infty
{d\nu\o2\pi}e^{it\nu}F\lf({\om-\nu\o2}\ri)F\lf({\om+\nu\o2}\ri)\ .
}
Note that $G(t)$ is an even-function of $t$.
The parameter $t$ can be interpreted physically as
the time difference between the bounce 
of the particle and the bounce of the hole.

In the following, we will set $\mu=0$ for simplicity
(its effects are easily restored).
Our expression \ANfinalform\ reproduces the well-known selection rule
that when $\mu=0$ 
the amplitude vanishes for $N={\rm even}$ \DemeterfiCM.
Below we will study the high energy behavior of the
non-vanishing $N={\rm odd}$ amplitude:
\eqn\ANodd{
{\cal A}_N=i^{N-1}\int_0^\infty{dt\o t}G(t)\prod_{j=1}^N2\sin\om_jt\ .
}

\subsec{Zeroth order approximation}

Suppose $G(t)$ is independent of $t$: $G(t)=G(0)$ $\forall t$.
Then the amplitude \ANodd\ becomes
\eqn\ANstep{\eqalign{
{\cal A}_N&=G(0)i^{-1}\int_0^\infty{dt\o t}\prod_{j=1}^N
(e^{it\om_j}-e^{-it\om_j}) \cr
&=G(0)\sum_{S}(-1)^{|S|}\int_0^\infty{dt\o t}\sin (\om_{\bar{S}}-\om_S)t\cr
&={\pi\o2}G(0)\sum_{S}(-1)^{|S|}\ep(\om_{\bar{S}}-\om_S) \cr
&=\pi G(0)\sum_{S}(-1)^{|S|}\th(\om-2\om_S)\ ,
}}
where $\ep(x)$ is the sign of $x$ and $\th(x)$ is the step function.
In the third equality of \ANstep, we used the formula 
$\int_0^\infty {dt\o t} \sin at
={\pi\o 2}\ep(a)$.
This expression \ANstep\ is the same 
as eq.(8.6) in \MooreZV, which was obtained by applying
the saddle point method directly to the integral \tilANshift.
As is obvious from the derivation of \ANstep,
this approximation does not capture an essential part of reflection,
contained in the $t$-dependence of $G(t)$.
We shall see shortly that $G(t)$ is approximately constant, but only
over a finite interval $t<t_0\sim O(\omega^{-1/2})$.  
It turns out that the integral
\ANodd\ has most of its support in the region $t>t_0$
when $N\gg\sqrt\omega$; a more detailed analysis is thus required.


\newsec{Saddle point approximation of reflection}
From the expression \ANodd, one can see that the amplitude 
vanishes linearly as $\om_i\riya0$, which is an example of the
low energy theorem for soft tachyon scattering \MooreZV.
Therefore, we expect that the dominant contribution is localized in the middle
of the phase space away from the boundary $\{\om_i=0\}$, and
when $N$ and $\om$ are sufficiently large
the amplitude ${\cal A}_N$ is sharply peaked around the mean value 
$\bra\om_i\ket={\om\o N}$.
This in turn implies that the saddle points of the $t$-integral are
located at $\sin\bra\om_i\ket t=\pm1$, {\it i.e.} 
$t=t_l={\pi(2l-1)\o 2\bra\om_i\ket}$ $(l=1,2,\cdots)$.
We checked numerically that the peak around 
the higher saddle points $t_{l>1}$ 
easily disappears when $\om_i$'s fluctuate around ${\om\o N}$, 
and thus we conclude that the first saddle
point $t_1={\pi N\o 2\om}$ gives the dominant contribution to the $t$-integral
\ANodd.

Numerical evaluation of $G(t)$ shows that it is nearly constant
for $t<t_0\sim O(\omega^{-1/2})$, and exponentially decaying
for $t>t_0$ (see figure \Gplot\ and below).  
This in turn leads to a change in
the characteristic behavior of the scattering amplitude
for $N\gg \sqrt\omega$ and $N\ll\sqrt\omega$;
we therefore analyze these two regimes separately.

\bigskip
{\vbox{{\epsfxsize=3.0in
        \nobreak
    \centerline{\epsfbox{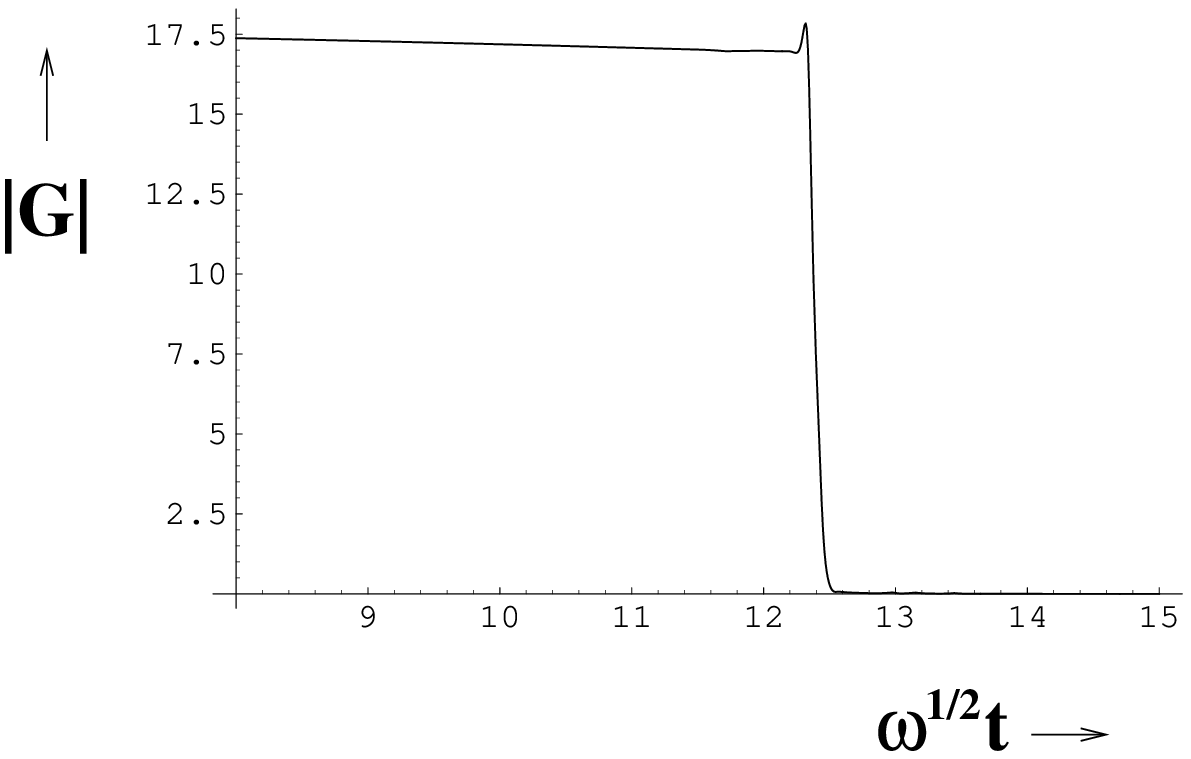}}
        \nobreak
    {\raggedright\it \vbox{
{\bf Figure \Gplot.}
{\it
Numerical evaluation of $|G(t)|$ for $\omega=400$, $q=0$.
The curve for $\omega=800$, $q=0$ is virtually identical
apart from a rescaling of the vertical scale by $\sqrt 2$.
} }}}}
\bigskip}

\subsec{$N\gg\rt{\om}$ case}
In the case $N\gg\rt{\om}$, the saddle point $t_1\sim{N\o\om}$ 
is much larger than 
the characteristic time scale $t_0\sim\om^{-\hf}$ of $G(t)$.
Therefore, 
the exponentially decaying behavior of $G(t)$ for $t\gg t_0$
is relavant for the evaluation of the $t$-integral \ANodd.
This behavior of $G(t)$
can be reproduced by 
closing the contour of the $\nu$ integral \GtFourier\
in the upper or lower half of the complex $\nu$-plane 
(depending on the sign of $t$), and
picking up the contribution from the poles 
of the $\Gamma$-functions in \Fxidef.
In this way $G(t)$ is estimated as
\eqn\Gassum{\eqalign{
G(t)&\sim 2\sum_{n=0}^\infty e^{-(1+q+i\om+2n)|t|}
{(-1)^n\o n!}{\Ga(1+q+i\om+n)\o\Ga(1+q+n)\Ga(-i\om-n)}\cr
&={2\sinh\pi\om\o\pi i}\sum_{n=0}^\infty e^{-(1+q+i\om+2n)|t|}
{1\o n!}{\Ga(1+q+i\om+n)\Ga(1+i\om+n)\o\Ga(1+q+n)}\cr
&={2\sinh\pi\om\o\pi i}e^{-(1+q+i\om)|t|}
{\Ga(1+q+i\om)\Ga(1+i\om)\o\Ga(1+q)}
{}_2F_1(1+q+i\om,1+i\om,1+q;e^{-2|t|})\ .
}}
We should emphasize that
this computation of closing the contour does not give an exact answer of $G(t)$
since the integrand of $G(t)$ does not decay as $\nu\riya\pm\infty$.

The $n=0$ term in the summation \Gassum\ gives
\eqn\Gexpt{
G(t)\sim {\Ga(1+q+i\om)\o\Ga(1+q)\Ga(-i\om)}e^{-(1+q+i\om)t}\ .
}
This term comes from poles in \GtFourier\ at
${\om\pm\nu}=i({1+q})$, 
where almost all the energy of the initial tachyon is carried 
by either the particle or by the hole. 
Including the higher order terms in the sum does not substantially
alter the results.

Now the $t$-integral in \ANodd\ can be evaluated by expanding the integrand 
around
the saddle point at
$t={\pi N\o2\om}$ and $\om_i={\om\o N}$
\eqn\tomexpand{
t={\pi N\o2\om}+u,\qquad \om_i={\om\o N}+\ep_i
	\qquad
\Big(\sum_{i=1}^N\ep_i=0,~|u|,|\ep_i|\ll1\Big)\ .
}
By performing the Gaussian integral over the fluctuation $u$,
${\cal A}_N$ is found to be
\eqn\ANapproxexp{\eqalign{
{\cal A}_N&\sim  {\Ga(1+q+i\om)\o\Ga(1+q)\Ga(-i\om)}2^N 
\int {du\o {\pi N\o2\om}}e^{-(1+q+i\om)({\pi N\o2\om}+u)}\exp\lf[
-{\om^2\o 2N}u^2
-\hf\lf({\pi N\o2\om}\ri)^2\sum_i\ep_i^2\ri] \cr
&\sim  {\Ga(1+q+i\om)\o\Ga(1+q)\Ga(-i\om)}2^Ne^{-(1+q+i\om){\pi N\o2\om}} 
{e^{-N/2}\o\rt{N}}  \exp\lf[
-\hf\lf({\pi N\o2\om}\ri)^2\sum_i\ep_i^2\ri]\ .
}}
Here we assumed $q\ll\om$.

\subsec{$N\ll\rt{\om}$ Case}
In the opposite case $1\ll N\ll \rt{\om}$,
the relevant behavior of $G(t)$ is the plateau region in $t\ll t_0$ 
(see figure 5). This behavior of $G(t)$ can be obtained by 
expanding the $\nu$-integral \GtFourier\ around $\nu=0$.
Physically, the point $\nu=0$ corresponds to the
situation where both particle and hole carry
the same amount of energy ${\om\o2}$. 
When $\nu$ is small, the integrand in \GtFourier\ is well approximated 
by a Taylor expansion in $\nu$
\eqn\Gtgauss{\eqalign{
G(t)&\sim \int{d\nu\o2\pi}F\lf({\om\o2}\ri)^2e^{it\nu+i{\om\o\om^2+q^2}\nu^2}\cr
&\sim \hf\rt{i(\om^2+q^2)\o\pi\om}F\lf({\om\o2}\ri)^2
e^{-i{\om^2+q^2\o4\om}t^2}
}}
with the saddle point
\eqn\nust{
\nu_*=-{\om^2+q^2\o2\om}t\ .
}
This approximation is self-consistent if $\nu_*$ is smaller than
the width of Gaussian in  \Gtgauss,
\eqn\nustbound{
|\nu_*| \sitarel{<}{\sim} \rt{\om^2+q^2\o\om}\quad \Rightarrow\quad
|t| \sitarel{<}{\sim} \rt{\om\o q^2+\om^2} \ .
}
Outside of this range, $G(t)$ is highly suppressed due to
the oscillatory factor $e^{it\nu}$. Therefore,
$G(t)$ is approximated as
\eqn\Gtapprox{
G(t)\sim \hf\rt{ i(\om^2+q^2)\o\pi\om}F\lf({\om\o2}\ri)^2
e^{-i{\om^2+q^2\o4\om}t^2}\th(t_0-|t|)\ ,
}
where
\eqn\tzero{
t_0=\ga\rt{\om\o q^2+\om^2}\ .
}
$\ga$ is a numerical coefficient independent of $q$ and $\om$.
Note that the time difference $t$ is very small
$t<\om^{-\hf}$ in the high energy limit.
Putting it all together, in this regime
the high energy $1\riya N$ amplitude is given by
\eqn\ANhigh{
{\cal A}_N=\hf i^{N-1}\rt{ i(\om^2+q^2)\o\pi\om}F\lf({\om\o2}\ri)^2
\int_0^{t_0}{dt\o t}
e^{-i{\om^2+q^2\o4\om}t^2}\prod_{j=1}^N2\sin\om_jt\ .
}
This approximation is consistent when ${N\o\om}\ll{1\o\rt{\om}}$; 
beyond this, the Taylor expansion \Gtgauss\
breaks down and it is better to use \Gassum.
Assuming $q\ll\om$, ${\cal A}_N$ becomes (up to a phase factor)
\eqn\ANsmallN{\eqalign{
{\cal A}_N&\sim \rt{\om}\int_0^{t_0}{dt\o t}
e^{-i{\om\o4}t^2}\prod_{j=1}^N2\sin\om_jt\cr
&\sim 2^N\rt{\om\o N}e^{-i{\om\o4}\lf({\pi N\o2\om}\ri)^2}\exp\lf[
-\hf\lf({\pi N\o2\om}\ri)^2\sum_i\ep_i^2\ri].
}}
Here we neglected the $u$-dependence coming from the
factor $e^{-i{\om\o4}t^2}$, since it is subleading when ${N\o\om}\ll1$.

\newsec{The number distribution at high energy}

One of the physically interesting quantities we can calculate is 
the number distribution $\PP_N$ of final state tachyons in the
$1\riya N$ scattering, which is defined by
\eqn\PNdef{
\PP_N={1\o N!}{1\o\om}\int_0^\infty\prod_{i=1}^N{d\om_i\o\om_i}\cob
\Bigl(\sum_{i}\om_i-\om\Bigr)\Bigl|{\cal A}_N(\om|\om_i)\Bigr|^2\ .
}
$\PP_N$ for the low energy and weak coupling scattering has been 
studied in \MooreZV\ and it was found that $\PP_N$ is almost 
a Poisson distribution.
We are interested in the high energy behavior of $\PP_N$.

If we substitute the naive step function approximation of ${\cal A}_N$
\ANstep\  
into the definition of $\PP_N$ \PNdef, it turns out that the 
resulting $\om_i$-integral can be performed exactly:
\eqn\PoddMPR{
\PP_{2k-1}\sim {(2\pi)^{2k-1}(2^{2k}-1)|B_{2k}|\o (2k)! \,\om}.
}
Here we used the approximation 
$|\pi G(0)|\sim \rt{\pi\om\o2}$ as in \MooreZV.%
\foot{Curiously, there is a nice answer for the generating function
of this result:
$$
  \sum_{k=1}^{\infty}\PP_{k}x^{k} = (2\omega)^{-1}\,  \tan (\pi x)\ .
$$
We also find it remarkable to discover yet another appearance
of the seemingly ubiquitous Bernoulli numbers.}
%
Note that \PoddMPR\ 
is a monotonically increasing function of $N=2k-1$.
It is argued \MooreZV\ that the approximation
\ANstep\ breaks down 
near the boundary of step functions.
However, one can argue that the contribution from the 
dangerous boundary region is negligible for small $N$,
so we expect \PoddMPR\ to be a good
approximation for sufficiently small $N$.

When $N$ becomes large, 
\PoddMPR\ cannot be trusted anymore
and we should use the approximation \ANapproxexp\ or \ANsmallN,
depending on the regime $N\gg\rt{\om}$ or $1\ll N\ll\rt{\om}$.
In both cases, the $\om_i$-integral can be approximated by
the Gaussian integral around the saddle point $\bra\om_i\ket={\om\o N}$
\eqn\omintapp{
{1\o\om}\int\prod_i{d\ep_i\o{\om\o N}}\cob\Bigl(\sum\ep_i\Bigr)\exp\lf[
-\lf({\pi N\o2\om}\ri)^2\sum_i\ep_i^2\ri]=\lf({2\o\rt{\pi}}\ri)^{N-1}
{N\o \om^2}\ .
}
Therefore, in the regime $N\gg\rt{\om}$, $\PP_N$ is given by
\eqn\PNlargeN{
\PP_N\sim {1\o N!}{\om^{2q}\o (q!)^2}\lf(8\o e\rt{\pi}\ri)^N 
e^{-(1+q){\pi N\o \om}}
}
and similarly in the regime $N\ll\rt{\om}$
\eqn\PNsmallN{
\PP_N\sim{1\o N!}{1\o\om}\lf({8\o\rt{\pi}}\ri)^N\ .
}
The discrepancy in these two formulae for $N\sim\sqrt\omega$
is subleading in $N$, and can be attributed to the different
approximations made in the two regimes.

\newsec{Discussion}

The main observation is that the probability to have
$N$ outgoing particles is a monotonically 
and steeply decreasing function of $N$, at least when
$N$ is sufficiently large.
On the other hand, \PoddMPR\ suggests the probability
grows with $N$ for $N$ small; indeed, in \MooreZV\
the probability to have more than one particle
in the final state at high energy was estimated 
by a unitarity argument%
\foot{Unitarity also tells us that 
the probability to send in a single initial tachyon
and then find $N$ tachyons in the final state,
is the same as that for the time reversal prosess of
$N\riya 1$ scattering.
Intuitively, it seems unlikely to have a single tachyon state
as the final state of $N$ tachyon scattering with large $N$, 
unless we fine-tune the initial
state.}
to be (noting that $\PP_1\sim {\pi\o2\om}$ according to \PoddMPR)
\eqn\Pnleqone{
\PP_{N>1}=1-\PP_1\sim 1-\frac{\pi}{2\omega}\ .
}
Taken together, our results
suggest that the distribution is peaked for some
reasonably small number of outgoing particles, larger than one.
Note that equation \PoddMPR\ gives $\PP_N\sim O(1)$
for $N\sim\log\omega$; we therefore expect $\PP_N$ 
to be peaked in this regime.

There is no characteristic feature for $N\sim\omega$ that
might indicate the presence of a black hole or similar object
as an intermediate state.  Instead, the configuration
that dominates high energy scattering is the motion of a single
particle-hole pair bouncing off the potential.  For $N\gg\sqrt\omega$,
the energy is carried predominantly by either the particle {\it or}
the hole; for $N\ll\sqrt\omega$, it is shared roughly equally by each.

A qualitative explanation for the rapid decrease of $\PP_N$
with $N$ comes from the fact that the $N$ tachyon state is
built from fermion bilinears $\psi^\dagger\psi$.
The initial state consists of one such bilinear, which 
then picks up a phase from reflection off the potential \zeroapotl.
The final state is a product of $N$ such bilinears, and the
overlap with the evolved initial state picks out the terms in
$\psi^\dagger\psi\cdots\psi^\dagger\psi$ where all but two
of the fermion operators act on one another.  This is a relatively
small fraction of the overall possibilities, since this operator
can create up to $N$ particle-hole pairs when acting on the vacuum.
The initial state selects only the part with one particle-hole pair.

\subsec{Comments on conservation laws}

The free fermionic character of the matrix model leads to
an infinite set of conserved quantities; the time-independent ones are
\eqn\conslaws{
  \QQ_\ell = \int d\vareps \;\vareps^\ell \,b^\dagger_\vareps b_\vareps^{~}
}
where $b, b^\dagger$ are fermion creation and annihilation operators.
Typically, black holes have no hair, 
suggesting that their conserved 
charges are $\QQ_1=\omega$, $\QQ_\ell=0$.%
\foot{Results to this effect were presented by A. Sen at
Strings2004 \senparis. See also 
\WittenFP.  }
States with these properties in the free fermion Hilbert space
consist of a macroscopic number of very soft tachyon excitations --
more or less a coherent state of soft tachyons.%
\foot{This feature is in accordance with higher dimensional
examples, where to form a macroscopic black hole requires
the excitation of a macroscopic number of branes.}
On the other hand, our one-particle initial state has
$\QQ_\ell\sim\omega^\ell$.  In order to form a black hole
as an intermediate state, an initial state consisting of a small
number of high energy tachyons would have to shed
these higher conserved charges, \eg\ by radiating a shell
of outgoing tachyons carrying them away, leaving behind
an intermediate state with the characteristics of the black hole.

The phase space available to soft particle-hole pairs is not large,
and the only objects with the charges $\QQ_\ell \sim \omega \delta_{\ell,0}$
are collections of soft fermion excitations of energies
$\{\omega_i\}$, $i=1,...,N$, relative to the filled Fermi sea;%
\foot{So that $\omega_i<0$ for holes and $\omega_i>0$ for particles.}
these yield the conserved charges
\eqn\softchg{
  \QQ_1=\sum_{i=1}^N \pm\omega_i=\omega\quad;\qquad 
  \QQ_\ell=\sum_{i=1}^N \pm\omega_i^\ell\sim o(N^{1-\ell})\ ,\quad\ell>1
}
with the sign $+$ for particles and $-$ for holes.
Since there is no
exponentially large phase space of black hole states,
the evolution is not drawn towards
the formation of an intermediate state with the 
conserved charges associated to the black hole --
there are not sufficiently many states to make visiting
such a configuration likely.

Now, to create an intermediate state corresponding to a finite
deformation of the Fermi sea, such as a blob of eigenvalues
separated away from the sea, one must consider a 
more general $N'\riya N$ scattering with $N',N\gg1$,
instead of the $1\riya N$ scattering studied in this paper.
This is because the number of particle-hole pairs appearing
in the $N'\riya N$ scattering
is bounded from above by
${\rm min}\{N',N\}$. In particular, 
only a single particle-hole pair contributes
in the $1\riya N$ scattering as we saw above,
and therefore the initial state we considered
has rather small overlap with the states carrying the
charges expected for a black hole state.
But if it happens that most high energy states
do not scatter into black holes, then this casts doubt
on the existence of any object that one should call a black hole.

The intuition gained from our study of $1\to N$
scattering shows why this soft-excitation state
will not have the properties of a black hole.
The Hawking radiation of a black hole consists predominantly
of quanta with energy of order the limiting temperature
$T_{\H}=\frac1{2\pi}\sqrt{\frac2\alp}$.  Such an outgoing state
would have to be constructed from an intermediate state
consisting of $N\gg\omega$ quanta, in order that the charges
\softchg\ approximate those of a black hole,
$\QQ_\ell=\omega\delta_{\ell,0}$;
again, the probability
that $N/\omega\equiv n\gg1$ soft particle-hole excitations will make
a Hawking quantum behaves like the overlap of a
state created by $(\psi^\dagger\psi)^n$ and one
created by $\psi^\dagger\psi$, which we expect to
be a rapidly decreasing function of $n$.

\subsec{Whither the 2D black hole?}  

The evidence seems to suggest that there
is no such object as the 2D black hole
in string theory.
The question then is why we were misled into thinking so
by the seemingly exact coset construction of a
worldsheet CFT describing strings propagating in a black hole background.%
\foot{We thank David Kutasov for the following argument;
see \refs{\KazakovPM,\AharonyXN}.}
Consider the Euclidean continuation of \twodbh, 
in which the putative Euclidean black hole
geometry is a capped semi-infinite
cylinder often referred to as the ``cigar''.
This geometry is strongly curved in the vicinity of the
tip of the cigar for the regime of interest, $k\sim 2$.
In fact, there is a conjectured strong-weak coupling duality
\refs{\GiveonPX,
\HoriAX,
\KazakovPM}
between the 2D black hole and Sine-Liouville theory,
\ie\ the tachyon background
\eqn\sinlio{
  T = \lambda\, e^{\frac1Q\phi}\cos \theta
}
describing a condensate of vortices,
where $\theta$ is the (axial) target Euclidean time coordinate,
and $Q=(k-2)^{-1/2}$.  The Euclidean cigar geometry is a weakly
coupled sigma model for $Q\to 0$, while the Sine-Liouville
background is weakly coupled for $Q\to\infty$.
The regime of interest for 2D black holes is the strongly
coupled one, where the Sine-Liouville description is more
appropriate, and one cannot use the cigar geometry
to extract reliable predictions.  
This observation is consistent with the fact
that the Hagedorn thermodyamics of high mass black holes is not
to be found in the matrix model.
The Sine-Liouville background appears to describe instead a classical
tachyon background of the 2D string, consistent with the
picture deduced from the consideration of conserved charges
in the previous subsection.

Perhaps the proper lesson to be drawn from all the known facts
is that high energy scattering is dominated by brane physics.
In higher dimensions, brane physics is black hole physics --
the degrees of freedom on brane intersections lead to 
a large degeneracy of black hole intermediate states,
which dominate the S-matrix.  
In two dimensions, the high energy scattering
is still dominated by branes -- the D0-branes that are the 
eigenvalue fermions of the matrix model -- but the kinematics
is so restrictive that they do not have a black hole density of states.
Instead of forming a black hole, a highly energetic tachyon
in a sense becomes a D0-brane that bounces high off the
background potential of figure \typezero, and then fragments
into a small number of very energetic tachyons in the final state.


\vskip 0.5cm
\noindent
{\bf Acknowledgments:~} We wish to thank David Kutasov
and Jongwon Park for useful discussions, 
and Sumit Das and Sergei Gukov for correspondence.
This work is supported in part by DOE grant DE-FG02-90ER40560.

\appendix{A}{The $2\riya N$ amplitude}

The $2\riya N$ amplitude is also written in terms
of the function $G(t)$ in \GtFourier.
When $\mu=0$ and $N={\rm even}$, the $2\riya N$
amplitude is given by
\eqn\AtwoN{\eqalign{
&{\cal A}_{2\riya N}\lf({\om\o2},{\om\o2}\Big|\,\om_i\ri)
={1\o\pi}\int_0^\infty{dt\o t}\int_0^\infty{ds\o s}\lf\{G_{\om}(t)
\lf[\prod_{l=1}^N2i\sin\om_l(s+t)-\prod_{l=1}^N
2i\sin\om_l(s-t)\ri]\ri. \cr
&\lf.-\sum_{A\cup B=\{1,\cdots N\}}G_{\om_A}*G_{\om_B}(t)
\lf[\prod_{l\in A}2i\sin\om_l(s+t)-\prod_{l\in A}2i\sin\om_l(s-t)\ri]
\prod_{j\in B}2i^{-1}\sin\om_js\ri\}\ .
}}
Here $f*g(t)$ is the convolution
\eqn\convol{
f*g(t)=\int_{-\infty}^\infty du f(u)g(t-u)\ .
}
Qualitatively, this amplitude will have a similar structure
to the $1\to N$ amplitude.  
Both $G$ and $G*G$ only have appreciable support for 
$t<O(\omega^{-1/2})$, and the sine factors are qualitatively
similar to the $1\to N$ result.  Thus we expect
qualitatively similar behavior, namely a monotonically decreasing
probability for finding $N$ particles in the final
state as a function of $N$.

\listrefs
\bye